\documentclass[9pt,twocolumn,twoside]{osajnl}

\journal{ol}

\setboolean{shortarticle}{true}
\usepackage[normalem]{ulem}
\usepackage{amsmath}
\usepackage{upgreek}

\title{Coherence transfer in an akinetic swept source OCT laser with optical feedback}

\author[1,2,3,$\dagger$*]{S. Slepneva}
\author[4,$\dagger$]{A. Kovalev}
\author[2,3]{N. Rebrova}
\author[4]{K. Grigorenko}
\author[4]{E. Viktorov}
\author[1]{G. Huyet}

\affil[1]{Universit\'e C\^ote d'Azur, CNRS, INPHYNI, Nice, France}
\affil[2]{Centre for Advanced Photonics and Process Analysis and Department of Physical Sciences, Cork Institute of Technology, Cork, Ireland}
\affil[3]{Tyndall National Institute, University College Cork, Cork, Ireland}
\affil[4]{ITMO University, Saint Petersburg, Russia}
\affil[$\dagger$]{These authors contributed equally to this Letter.}

\affil[*]{Corresponding author: svetlana.slepneva@inphyni.cnrs.fr}

\begin{abstract}
We theoretically investigate the influence of optical feedback onto the dynamics of a semiconductor swept source laser. In particular, we show that optical feedback can be used to lock the phase of the successive lasing modes of a multi-section semiconductor laser commonly used for Optical Coherence Tomography (OCT) applications.  We also identify two different regimes called sliding frequency self-mixing and sliding frequency mode-locking. The second regime demonstrates sub-nanosecond sliding frequency pulses for non-linear optics applications. 
\end{abstract}

\setboolean{displaycopyright}{true} 

\begin{document}

\maketitle

 The development of Swept Source Optical Coherence Tomography (SS-OCT) technique ~\cite{Chinn1997} has increased the demand for ultra-fast, broadband and highly coherent swept sources~\cite{klein2017high}. Swept sources are lasers that optically sweep over a wide spectrum range while displaying a small intensity modulation. The quality of OCT images depends on various parameters of the light sources including the spectral width, the relative intensity noise and the coherence length. This last parameter, which is inversely proportional to the instantaneous linewidth, is the key in determining the imaging depth. While frequency swept sources originally relied on external cavity configuration including an intra-cavity tunable element such as a rotating mirror~\cite{johnson2015analysis}, state-of-the-art sources include long cavity Fourier Domain Mode-Locked (FDML) Lasers~\cite{huber2006fourier}, micro-electro-mechanical system (MEMS) based Vertical Cavity Surface Emitting Lasers (VCSELs)~\cite{john2015wideband}, short cavity laser with MEMS tunable Fabry-P\'erot filter~\cite{bart2017,bart2018} and tunable Sampled Grating Distributed Bragg Reflector  (SG-DBR) Lasers~\cite{bonesi2014akinetic}. While the first three technologies have demonstrated long coherence length that enable long imaging depth, SG-DBR lasers present the advantage of not including any mechanically movable parts. However, numerous mode hops across a single sweep deteriorate their coherence properties. 

This paper aims to investigate a means to improve the coherence length of akinetic SG-DBR lasers by locking the successive longitudinal sweeping modes. Our study targets the laser configuration based on akinetic multi-section technology, similar to that presented in~\cite{bonesi2014akinetic}. The advantage of this swept source is its stability, high speed and potential to be integrated into a compact device.  Such lasers have been used as widely tunable lasers in the telecom industry for many years~\cite{jayaraman1993theory} and have, more recently, been considered for OCT applications~\cite{salas2018compact}. The operation of these devices relies on the Vernier effect between two sample gratings that leads to the single mode operation~\cite{jayaraman1993theory}. By varying the injection currents in these gratings, it is possible to sequentially select and tune each single mode over several nanometers. Wide bandwidth tuning is obtained with a subsequent tuning of a sequence of Vernier modes, thus the laser output is a sequence of highly coherent single mode sweeps separated by transients. The coherence length of each mode could be estimated using the Schawlow-Townes approach \cite{Schawlow1958} taking into account various noise sources. While each mode has similar instantaneous linewidth, the hops between successive Vernier modes lead to random optical phase shifts that deteriorate the coherence of these lasers.

To improve the coherence, we use external optical feedback technique. External optical feedback has been used as an effective tool to improve the coherence of continuous wave (cw) or tunable lasers, though under some parameters it can lead to the coherence deterioration~\cite{van1995semiconductor}. Here we introduce the concept of optical feedback into a fast-sweeping laser using an example of akinetic configuration, and derive the locking conditions for the subsequent modes.
While in an akinetic laser each subsequent single mode starts lasing from spontaneous emission, our method implies optical feedback to seed each lasing mode by a returned portion of the preceding lasing mode. In analogy with cw lasers, it is possible to derive Adler-like equation describing the locking conditions \cite{adler1946}. By introducing optical feedback, we show that it is possible to lock the phase of successive modes, and and maintain coherence over the mode hops.

\begin{figure}[tb]
\centering
\includegraphics[width=0.55\linewidth]{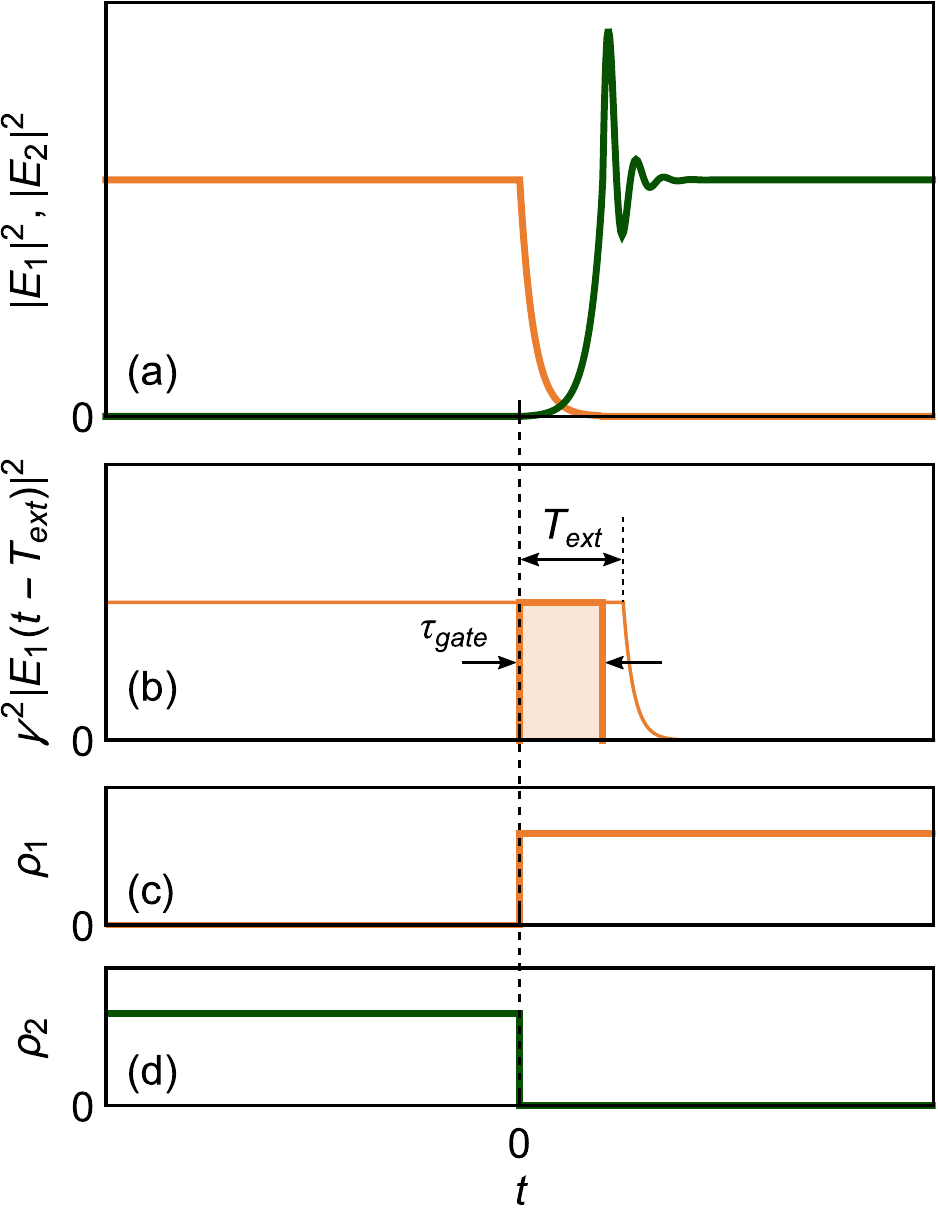}
\caption{The time-gated feedback. (a) A schematic illustration (both axes are in arbitrary units) of a single mode hop event happening at time 0 between mode 1 (orange) and mode 2 (green). We define $t=0$ as the moment when mode 1 starts to decay and mode 2 starts to build up the lasing via relaxation oscillations. (b) Intensity of the time-delayed attenuated mode 1 given by $\gamma^2 |E_1(t-T_{ext})|^2$ (thin line) and the box highlighting the period $\tau_{gate}$ when the gate is open (set by the modulator), and described by $\eta(t)^2 |E_1(t-T_{ext})|^2$. During this period, the part of mode 1, reflected from the external mirror, is reinjected into the laser to seed mode 2. (c) and (d) visualise the switch of the values of the functions $\rho_1(t)$ and $\rho_2(t)$, correspondingly (see  Eqs~\ref{e1-gated}-\ref{e2-gated}). Note, the ratio $\tau_{gate}/T_{ext}$ can be adjusted.  }
\label{fig:modehop}
\end{figure}

\begin{figure}[tb]
\centering
\includegraphics[width=0.6\linewidth]{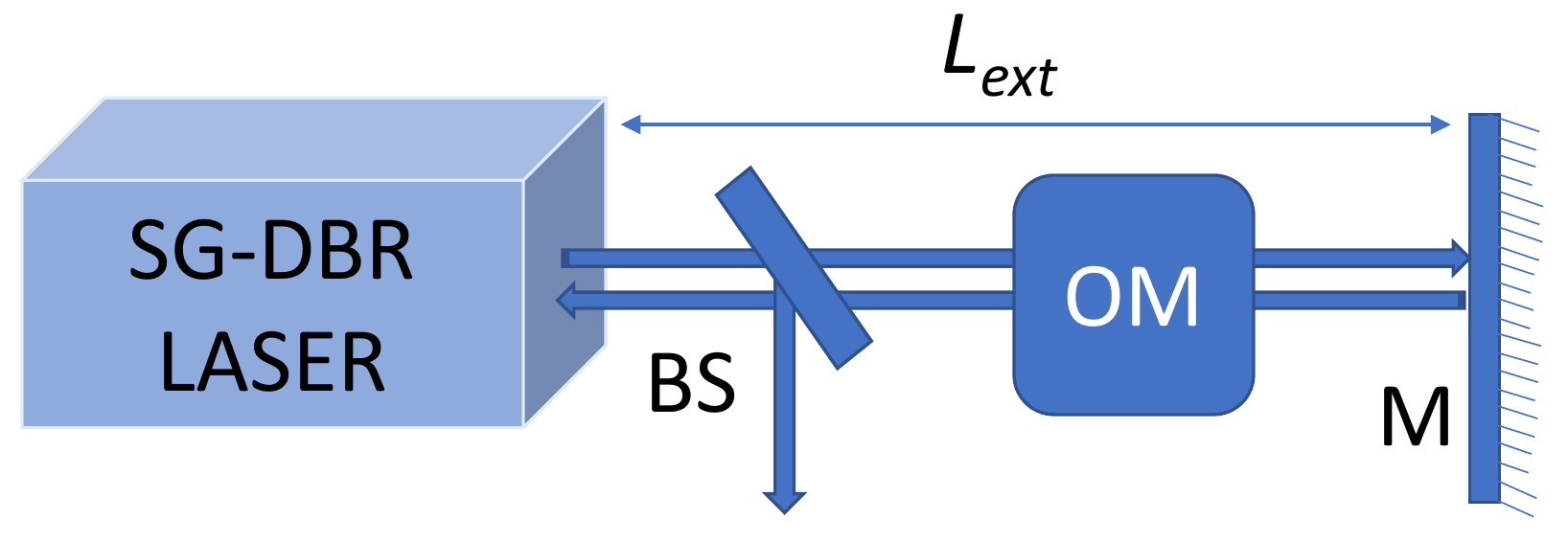}
\caption{A schematic setup of the feedback. M - backreflecting mirror; BS - beam splitter; OM - optical modulator provides periodically opening window for the gated feedback; $L_{ext}$ corresponds to the optical path length of the external cavity and defines the delay time $T_{ext}=\frac{2L_{ext}}{c}$, where $c$ is the speed of light in the cavity. }
\label{fig:setup}
\end{figure}

Let us consider the mode hopping event between the two subsequent modes~1 and~2 with complex amplitudes $E_1(t)$ and $E_2(t)$ correspondingly. The $1^{st}$ mode is switched on for $t\leq0$ and the $2^{nd}$ mode is on for $t>0$.  First, we consider the time-gated synchronisation of two successive modes, where the feedback is turned on just before the mode hop as shown in Fig.~\ref{fig:modehop}(a). 
The part of the sweep emitted by mode $1$ of duration $\tau_{gate}$, just before this mode is switched off,  is stored in an external cavity of length $L_{ext}$ in order to seed the second mode as illustrated in Fig.~\ref{fig:modehop}(a). Experimentally, that will require additional components in the external cavity, such as an optical amplitude modulator or a MEMS-actuated mirror, in order to control the feedback duration (Fig.~\ref{fig:setup}).  

The evolution of the complex field amplitudes $E_{1,2}(t)$ and carrier density $N(t)$ of the SG-DBR laser can be described by the following set of equations, with the time $t$ normalised by the free running laser photon lifetime:
\begin{equation*}
\begin{aligned}
\dot E_1(t)&=&-\rho_1 (t) E_1(t) + i\Omega_1(t)  E_1(t) +
\end{aligned}
\end{equation*}
\begin{equation}
\begin{aligned}
\frac{1}{2}\left( 1+ i\alpha \right) \left( N(t)-1\right)  E_1(t),\label{e1-gated}\\
\end{aligned}
\end{equation}
\begin{equation*}
\begin{aligned}
\dot E_2(t)&=& -\rho_2 (t) E_2(t) + i\Omega_2(t)  E_2(t) +\\
\end{aligned}
\end{equation*}
\begin{equation}
\begin{aligned}
\frac{1}{2}\left( 1+ i\alpha \right) \left( N(t)-1\right)  E_2(t) + \eta (t) E_1(t-T_{ext}),\label{e2-gated}\\
\end{aligned}
\end{equation}
\begin{equation}
\begin{aligned}
\dot N(t)&=&-\gamma_N\left( N(t)-g_0 +N(t) \left(|E_1(t)|^2+|E_2(t)|^2\right)\right),\label{carrier-hop}
\end{aligned}
\end{equation}
where the electric field of each mode is $E_n(t)=R_n\exp({i\phi_n(t)})$ with, here and below, $n=1,2$ staying for mode 1 or 2 respectively; dot corresponds to differentiation with respect to normalized time $t$; $R_n$ is the amplitude of the electric field and $\phi_n(t)=\Omega_n(t)+\xi_n$ is the phase, where 
$\Omega_n(t)=\omega_{0n} + a t$ is the angular frequency of each mode, which describes the linear sweep starting from $\omega_{0n}$ at a sweeping rate of $a$, $\xi_n$ is arbitrary initial phase; $\alpha$ is the linewidth enhancement factor; $\gamma_N$ is the carrier decay rate and $g_0$ is the pump parameter. The functions $\rho_{1,2}(t)$ describe the Vernier effect; $\rho_1(t)= \kappa_1 \Theta (t)$, where $\Theta(t)$ is the Heaviside step function and $\kappa_1$ is the decay rate of mode 1 when this mode becomes strongly damped as schematically explained in Fig.~\ref{fig:modehop} (b).   Similarly, $\rho_2(t)=\kappa_2 \Theta (-t)$ (Fig.~\ref{fig:modehop}(c)). The term $\eta (t) E_1(t-T_{ext})$ describes the time-gated optical feedback from mode 1 into mode 2, where $\eta (t) = \gamma \Pi(t)$ is the time varying feedback level: $\gamma$ is the feedback strength, $\Pi(t)=\Theta(t)-\Theta(t-\tau_{gate})$ is the rectangular function with duration $\tau_{gate}$; $T_{ext}$ corresponds to the time delay of the light of mode 1 reinjected into mode 2. 

\begin{figure}[htbp]
\centering
\includegraphics[width=0.4\textwidth]{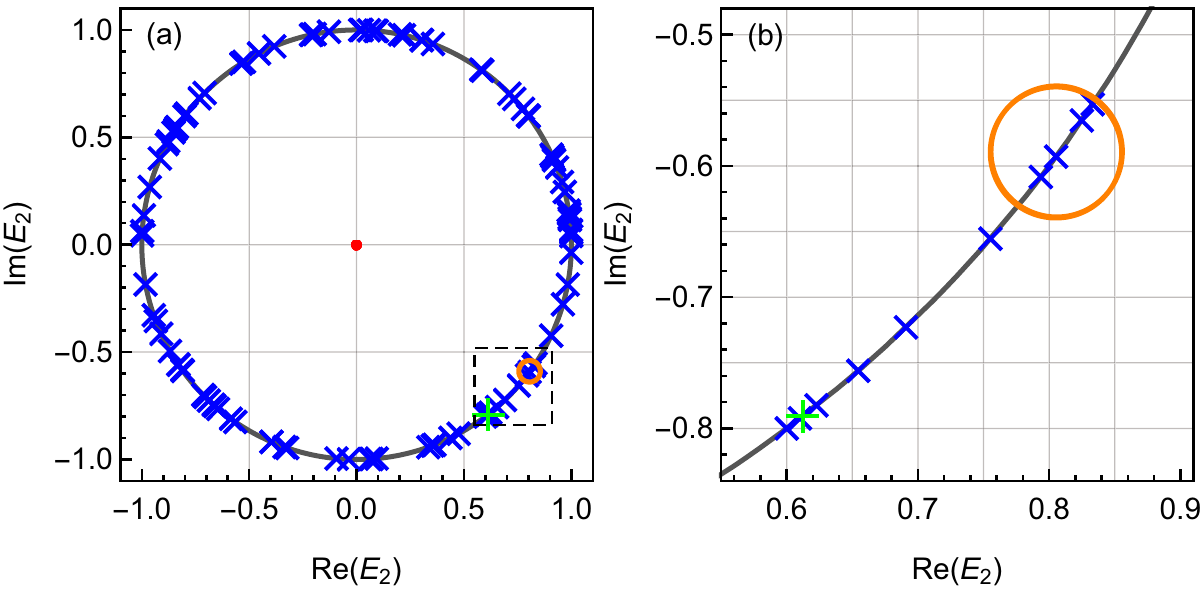} 
\caption{(a) Real and imaginary parts of the electric field of mode 2 after $t=20000$ calculated for 100 different initial conditions for three cases 1) free-running (the grey circle corresponding to the randomly distributed phase and the blue crosses indicate 100 calculated solutions); 2) time-gated feedback (green cross indicating the locking of the phase of mode 2 to mode 1); and 3) continuous feedback (the orange circle indicating the limit cycle). The red dot near the origin accumulates 100 different initial conditions of the electric field of mode 2, the initial conditions of mode 1 remains the same. (b) Zoom of the area in the dash square in (a). The parameters are $T_{ext}=200$, $\gamma_N=0.001$, $g_0=2$, $\alpha=5$, $\kappa_1=\kappa_2=1$, $\gamma=0.2$, $a = -0.002$. For the simulation $\tau_{gate}=T_{ext}$.}
\label{fig:field}
\end{figure}

Without optical feedback, {\it i.e.} if $\gamma =0$, the temporal evolution of the optical phase of the two modes are independent since they are not coupled and, as a result, the laser loses its coherence at each mode hop. To illustrate this point, we numerically integrated Eqs.~(\ref{e1-gated})-(\ref{carrier-hop})  for $100$ different initial phase values (keeping the same value for the amplitude). In this case, the phase of the electric field in the complex plane for time $t=20000$ is randomly distributed on a circle with radius $\sqrt{g_0-1}$, as shown in Fig.~\ref{fig:field}(a), thus demonstrating the loss of coherence at each mode hop. When the time-gated feedback is applied to the same situation, all the transients of mode~$2$ converge towards the same value of the electric field which is locked to mode~1 as shown by the green cross in Fig.~\ref{fig:field}(a) (and zoomed in Fig.~\ref{fig:field}(b)), demonstrating the coherence transfer from mode 1 to mode 2.  

To gain further insight on the locking dynamics, we can assume that the intensities of the mode $|E_2(t)|^2$ and $|E_1(t-T_{ext})|^2$ are constant and derive an equation for the phase difference $\delta\phi(t) =\phi_2(t)-\phi_1(t)$:
\begin{eqnarray}
 \delta\dot\phi (t)&=&  \delta\Omega(t)  - \gamma \frac{R_1}{R_2}\sqrt{1+\alpha^2}\sin\delta\phi(t),
\end{eqnarray}
where $\delta\Omega(t)=\Omega_2 (t) - \Omega_1 (t-T_{ext} )$. The two modes can only lock if the frequency difference between them follows the inequality
\begin{eqnarray}
|\delta\Omega(t)| &<&   \gamma \frac{R_1}{R_2}\sqrt{1+\alpha^2}.
\end{eqnarray}

In order to retain the constant phase between the successive modes, it is important that the timing jitter of the modulator does not cause phase fluctuations. To address this point, we should note that the phase difference $\delta\phi$ is constant when the two modes are locked, and their angular frequency difference is $\delta\Omega$. If the switching time of the modulator is shorter than this frequency beating, then the phase fluctuations will be insignificant. 

While the gated optical feedback can lock two consecutive modes, experimentally, it requires the incorporation of an optical modulator in the external cavity to switch on and off the optical feedback and, as an alternative, we consider the possibility of having continuous optical feedback in an SG-DBR laser. 
The effect of optical feedback on the dynamics of semiconductor lasers has been the subject of numerous studies as it may either strongly improve or deteriorate the coherence properties of the laser depending on the parameters. 
In the case of a wavelength sweeping laser, the light fed back into the cavity is frequency shifted by $\Delta=aT_{ext}$ with the instantaneous laser frequency $\dot\phi_{n}(t)$. The parameter $\Delta$ corresponds to the frequency shift between the light in the cavity and the light reinjected into the cavity as a result of the feedback. As a result, the effect of optical feedback in a frequency swept source is closely related to the dynamics of frequency shifted feedback previously investigated in~\cite{yatsenko2004theory,Yun1997} but the frequency shift can be of the order of a few GHz and, as a result, can be much larger than that obtained using an external modulator. In the case of a swept source,  the main difference between using continuous instead of time-gated feedback is that continuous feedback may destabilise the laser, and deteriorate the coherence of each mode. 

To investigate this, we consider the dynamics of a single mode frequency swept source with constant continuous optical feedback. The equations describing the evolution of this mode read

\begin{equation*}
\begin{aligned}
\dot E(t) &=&i\Omega \left( t\right) E(t)+\frac{1}{2}\left( 1+ i\alpha \right) \left( N(t)- 1 \right) E(t) + 
\end{aligned}
 \end{equation*}
\begin{equation}
\begin{aligned}
+\gamma E\left( t- T_{ext}\right),\label{delayfield1}\\
 \end{aligned}
 \end{equation}
 \begin{eqnarray}
\begin{aligned}
\dot N (t)&=& -\gamma_N \left( N-g_0(t) + N(t)|E(t)|^2 \right).\label{delaycarrier1}
\end{aligned}
\end{eqnarray}
The frequency $\Omega(t)$ is assumed to be increasing linearly with time, {\it i.e.} $\Omega(t)= \omega + a t$. For $a=0$, these equations reduce to the well known Lang-Kobayashi equations \cite{Lang1980}. In this regime, the external cavity modes are given by $E(t)=R \exp (i\omega t)$, where $\omega$ is a solution of
\begin{eqnarray}
\omega = -\gamma \sqrt{ 1+ \alpha^2}\sin\left( \omega T_{ext} +\arctan \alpha \right).
\end{eqnarray}

From this equation, we note that the external cavity modes (ECMs) lie within the frequency range defined by $|\omega | <\gamma \sqrt{1+\alpha^2}.$ Since these modes may be stable or unstable depending on various parameters, the laser can potentially demonstrate a wide range of behaviour including narrow linewidth emission, mode-hopping, undamped relaxation oscillations or low frequency fluctuations.
\begin{figure}[tb]
\centering
\includegraphics[width=0.75\linewidth]{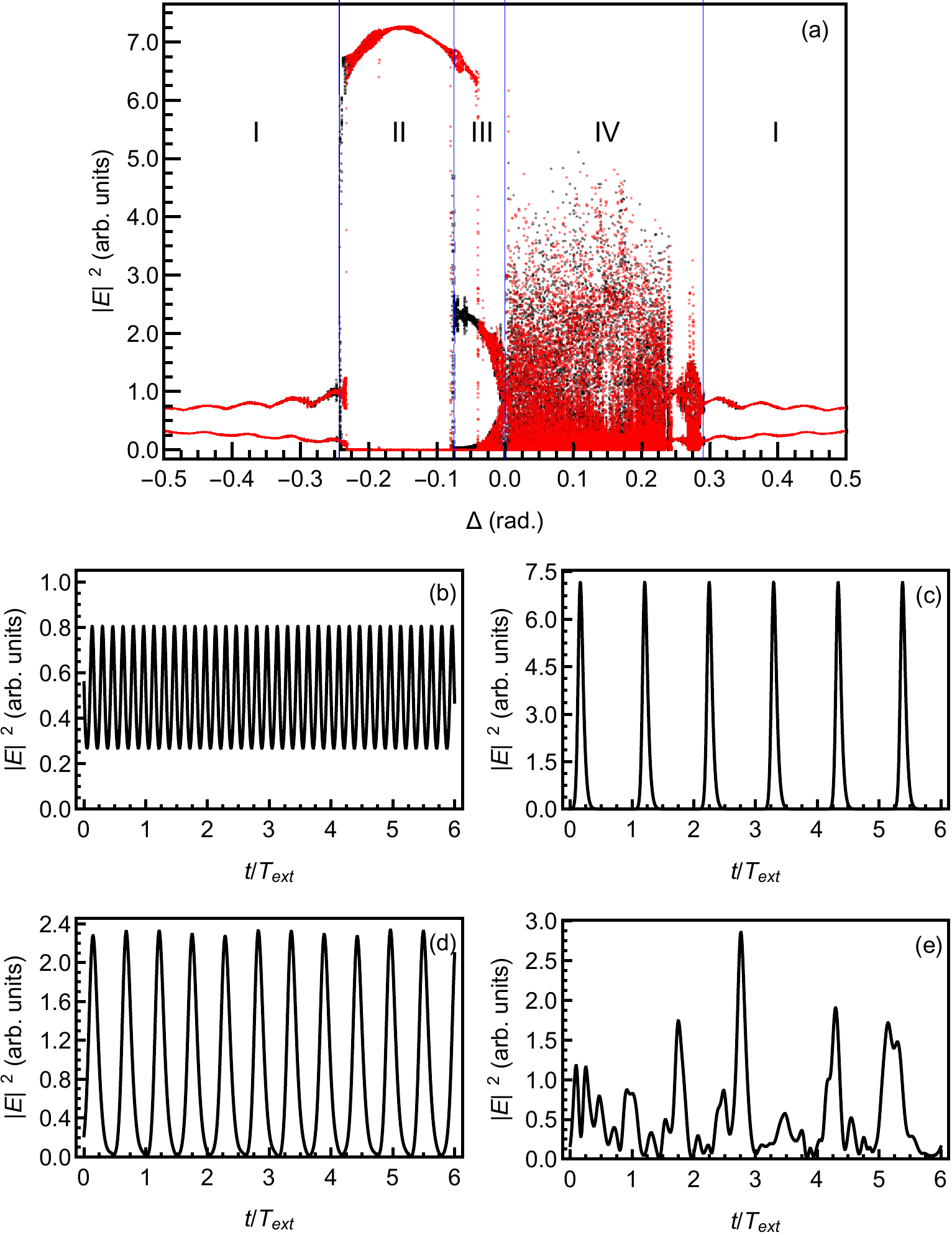}
\caption{(a) Bifurcation diagram of the swept source laser with continuous optical feedback demonstrating extrema of the intensity. The black (red) color corresponds to the decrease (increase) of the parameter $\Delta$. The numbers I-IV denote the dynamical regimes, and corresponding intensity time traces are given for sliding frequency self-mixing regime I at $\Delta = 0.4$ (b), sliding frequency mode-locking regime II at $\Delta = -0.12$ (c), second harmonic mode-locked regime III at $\Delta = -0.06$ (d), chaotic IV at $\Delta = 0.2$ (e). The other parameters: $T_{ext} = 100$, $\gamma_N = 0.01$, $g_0 = 1.5$, $\alpha = 2$, $\gamma = 0.1$.}
\label{fig:bifur}
\end{figure}

In order to investigate various dynamical regimes that can be demonstrated by a swept source with constant feedback, we numerically integrated equations (\ref{delayfield1})~and~(\ref{delaycarrier1}), and performed bifurcation analysis which revealed four possible outputs depending on parameter $\Delta$ which varies from $-0.5$ to $0.5$ rad. Figure~\ref{fig:bifur}(a) shows the bifurcation diagram, and Figs.~\ref{fig:bifur}(b)-\ref{fig:bifur}(e) demonstrate simulated intensity time traces to illustrate each regime. At large values of $|\Delta|$, the laser operates in self-mixing regime \cite{rudd1968laser}, where the temporal evolution of the electric field displays periodic oscillations with frequency corresponding to the frequency beating $\Delta$ between the emitted and re-injected light (see Region I in Fig.~\ref{fig:bifur}(a) and Fig.~\ref{fig:bifur}(b)). As $\Delta$ increases, the lasers enters regime II and emits periodic pulses corresponding to sliding frequency mode-locking (Fig.~\ref{fig:bifur}(c)) similar to those observed in~\cite{bart2017,Butler2019}. For higher negative values of $\Delta$, the laser reachers region III  which is similar to region II, but with the doubled frequency (Fig.~\ref{fig:bifur}(d)). For $\Delta=0$ and for small positive values of $\Delta$ the laser enters region IV where the above equations display a chaotic output characteristic of instabilities commonly observed in the Lang-Kobayashi equations~\cite{van1995semiconductor} (Fig.~\ref{fig:bifur}(e)). 

The observed regimes I, II and III are of high interest for non-linear optics applications including various swept-source based imaging modalities such as hyperspectral imaging \cite{begin2011} or SS-OCT. The lasers based on MEMS have been used for the latter technique~\cite{bart2017}. Such devices operate in a pulsed regime with the repetition period corresponding to the cavity roundtrip time or half of the cavity roundtrip time, depending on the sweeping rate \cite{avrutin2019,Butler2019}. The pulses are generated via the coherence transfer between the successive modes due to four-wave mixing as an interplay between the sweeping rate and the transmission bandwidth of the tunable filter. This results in undesired increase of the laser linewidth with the increase of the sweeping rate~\cite{Slepneva2014}, however, the advantage of this laser is the possibility to  implement a coherence revival scheme \cite{dhalla2012} for application in imaging. The frequency sweeping pulses can also be generated from mode-locked lasers~\cite{begin2011} or FDML lasers applying the pulse compression~\cite{eigenwillig2013} or master oscillator power amplifier-based~\cite{karpf2019fourier} methods. If compared to these methods of generation short pulses with sweeping frequency, the continuous optical feedback can be used for generation of different adjustable regimes, particularly, a train of high-intensity sub-ns pulses with frequency variation between the pulses equal to $\Delta$, which can be adjusted by changing either the sweeping rate $a$ or the delay time $T_{ext}$. 

\begin{figure}[htbp]
\centering
\includegraphics[width=0.6\linewidth]{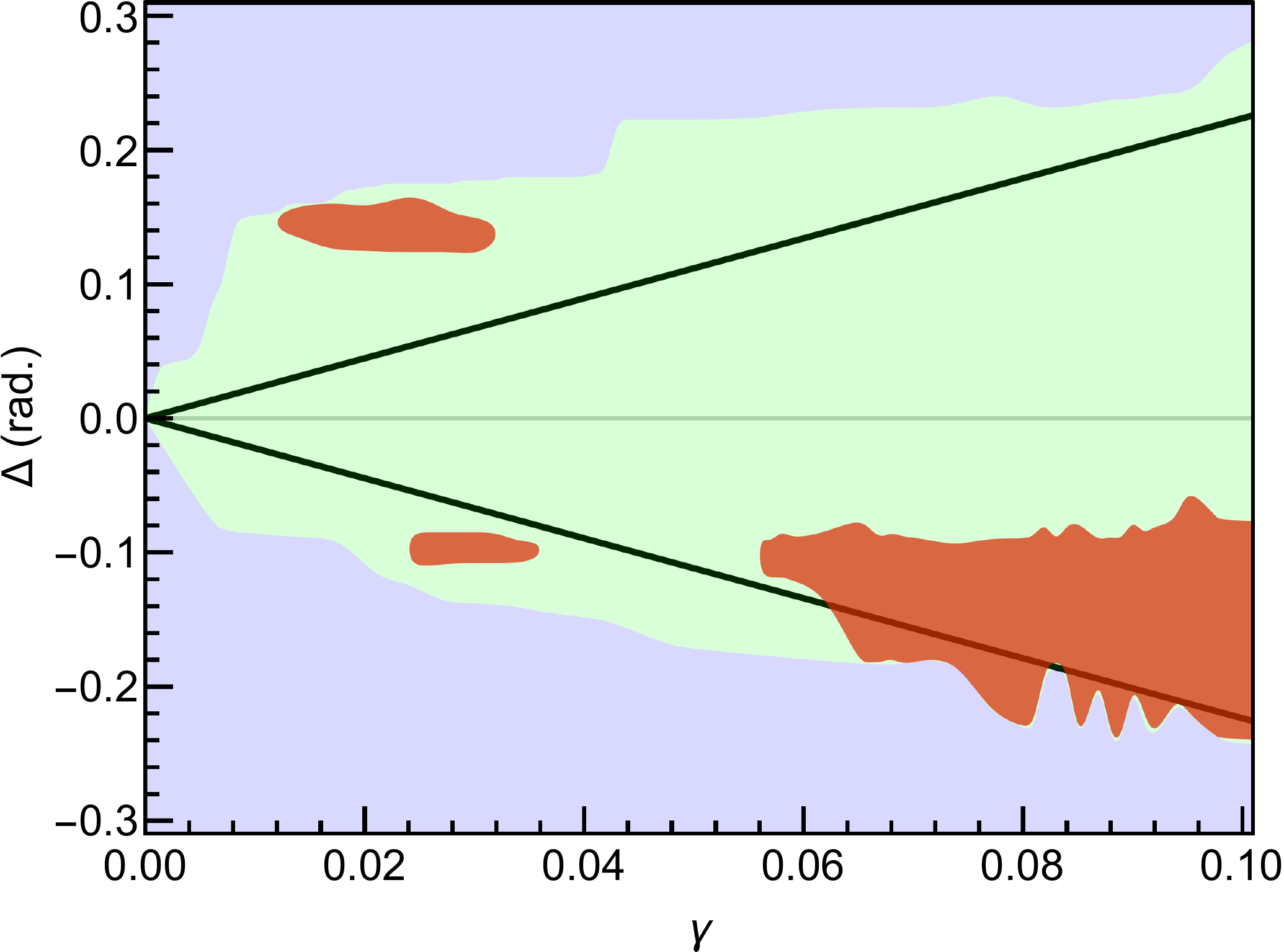}
\caption{The two-dimensional map of the dynamical regimes for the continuous feedback swept source. The black lines correspond to $|\Delta|=\gamma \sqrt{1+\alpha^2}$ defining the borders of the region within which the ECMs exist. The light green area indicates non-periodic output (regimes III and IV), the purple indicates sliding frequency self-mixing regime (regime I), and the red areas indicate sliding frequency mode locking (regime II). The other parameters are the same as in Fig.~\ref{fig:bifur}.}
\label{fig:cone}
\end{figure}
The lasing regimes of the SG-DBR laser under the continuous feedback can be controlled by varying the frequency shift $\Delta$ and feedback strength $\gamma$,  as demonstrated in the the two-dimensional map in Fig.~\ref{fig:cone}. The black lines correspond to the relation $|\Delta|=\gamma \sqrt{1+\alpha^2}$ indicating the frequency range within which the ECMs exist. Outside of this range, for $|\Delta|> \gamma \sqrt{1+ \alpha^2}$, the purple area corresponds to the sliding frequency self-mixing regime I, where the laser does not lock with the re-injected light. The green area corresponds to the non-periodic regimes III and IV and the localised red areas indicate the appearance of sliding frequency mode-locked regime II.
Interestingly, as the major sliding frequency mode-locked regime can be obtained for the range of the feedback strength from 0.056 to 0.099  and negative frequency shifts, the analysis reveals the appearance of two additional small areas  for the both positive and negative frequency shift values and smaller feedback strength.

In conclusion, we have shown that it is possible to lock the phase of successive modes of a multi-section swept source semiconductor laser using optical feedback. As a result of locking, the coherence of the laser does not deteriorate after the mode hops. In addition, we have demonstrated that optical feedback can lead to the appearance of sliding frequency self-mixing and sliding-frequency mode locking similar to that observed in short cavity swept sources laser, which offers an opportunity of exploiting schemes such as coherence revival \cite{dhalla2012}. The generation of sliding frequency mode locking delivering subnanosecond pulses is attractive for nonlinear optics applications \cite{begin2011}.
While the power of the proposed technique has been demonstrated on the model of the multi-section semiconductor lasers, it can also be adapted to other swept source configurations exhibiting mode hops, or serve as a tool to couple several swept sources to increase the sweeping range with maintained coherence. 

\section*{Funding}
H2020-MSCA-IF-2017, ICOFAS project (800290); H2020-MSCA-RISE-2018 HALT; OPTIMAL project granted by the European Union by means of the Fond Européen de développement regional, FEDER; Government of Russian Federation (08-08).

\bibliography{sample}

\bibliographyfullrefs{sample}

\end{document}